\documentclass{aa}
\usepackage[varg]{txfonts}
\usepackage[flushleft]{threeparttable}
\usepackage{graphicx}
\usepackage{natbib}
\usepackage{isotope}
\usepackage{epstopdf}
\usepackage{epsfig}
\usepackage{rotating}
\usepackage{amsmath}
\bibpunct{(}{)}{;}{a}{}{,} 

\makeatletter
\newcommand*{\rom}[1]{\expandafter\@slowromancap\romannumeral #1@}
\makeatother

\begin{document} 

\authorrunning{Wu et al. 2018}
\titlerunning{Solar Total and Spectral Irradiance Reconstruction Over the Last 9000 Years}

\title{Solar total and spectral irradiance reconstruction over the last 9000 years}
\author{C.-J. Wu\inst{1,2}
 \and
 N. A. Krivova\inst{1}
 \and
 S. K. Solanki\inst{1,3}
 \and
 I. G. Usoskin\inst{4}
}

\institute{Max-Planck-Institut f\"ur Sonnensystemforschung, Justus-von-Liebig-Weg 3, G{\"o}ttingen, Germany\\
  \email{wu@mps.mpg.de}
  \and
  Georg-August-Universit{\"a}t G{\"o}ttingen, Institute for Astrophysics, G{\"o}ttingen, Germany
  \and
  School of Space Research, Kyung Hee University, Yongin, South Korea
  \and
  Space Climate Research Unit and Sodankyl\"a Geophysical Observatory, University of Oulu, Finland
}


 
\abstract
   {Changes in solar irradiance and in its spectral distribution are among the main natural drivers of the climate on Earth. However, irradiance measurements are only available for less than four decades, while assessment of solar influence on Earth requires much longer records.}
   {The aim of this work is to provide the most up-to-date physics-based reconstruction of the solar total and spectral irradiance (TSI/SSI) over the last nine millennia.}
   {The concentrations of the cosmogenic isotopes \isotope[14][]C and \isotope[10][]Be in natural archives have been converted to decadally averaged sunspot numbers through a chain of physics-based models. TSI and SSI are reconstructed with an updated SATIRE model. Reconstructions are carried out for each isotope record separately, as well as for their composite.}
   {We present the first ever SSI reconstruction over the last 9000 years from the individual \isotope[14][]C and \isotope[10][]Be records as well as from their newest composite. The reconstruction employs physics-based models to describe the involved processes at each step of the procedure. }
   {Irradiance reconstructions based on two different cosmogenic isotope records, those of \isotope[14][]C and \isotope[10][]Be, agree well with each other in their long-term trends despite their different geochemical paths in the atmosphere of 
Earth. Over the last 9000 years, the reconstructed secular variability in TSI is of the order of 0.11\%, or 1.5 W/m$^2$. After the Maunder minimum, the reconstruction from the cosmogenic isotopes is consistent with that from the direct sunspot number observation. Furthermore, over the nineteenth century, the agreement of irradiance reconstructions using isotope records with the reconstruction from the sunspot number by \cite{Chatzistergos17} is better than that with the reconstruction from the WDC-SILSO series \citep{Clette14}, with a lower $\chi^2$-value.}


   \maketitle


\section{Introduction}
\label{sect:paper2:intro}

The Sun is the dominant external energy source of the Earth \citep{Kren17} and thus even small changes in irradiation from the Sun are expected to affect the Earth's climate \citep{Hansen2000, Haigh01, Haigh03, Gray10, Solanki13}. Although solar variability and its influence on Earth have been studied for a long time using various approaches, the physical processes and the level of solar influence on the coupled atmosphere-ocean system of the Earth are not yet fully understood. Variability in solar irradiance is considered to be among the possible mechanisms of this latter influence. The total solar irradiance (TSI), which is dominated by radiation in the visible and infrared (IR)
 bands of the spectrum, is the spectrally integrated energy flux per unit area normalised to 1 AU, while the solar spectral irradiance (SSI) spectral range, which includes ultraviolet (UV),  is the flux per unit wavelength. These two categories of irradiance, TSI and SSI, form the main agents of the so-called hypothesized bottom-up and top-down mechanisms, respectively \citep{Gray10}. The former mechanism is induced by the solar irradiance absorbed by Earth's surface while the latter is associated with the interaction between the UV irradiance and the stratosphere \citep{Haigh94, Haigh01}. As UV radiation below 400 nm most probably contributes more than half of the TSI variability \citep{Krivova06, Ermolli13, Yeo14, Yeo17, Morrill14, Woods15}, SSI and the UV range are of particular interest for studies of solar influence on climate. 

To better understand the mechanisms of solar impact on climate, long-term records of solar irradiance are required. Both TSI and SSI have been measured since 1978 using space-based instruments, yet for a longer timescale suitable models are needed. A number of models have been published \citep{Lean2000, Krivova07, Krivova10, Shapiro10, DasiEspuig14, DasiEspuig16, Coddington16} that reconstruct TSI and SSI on decadal to centennial timescales from various solar activity proxies, such as the Ca{\small \rom{2}} K index, solar radio flux f10.7cm, sunspot areas, or the sunspot number. Isotope concentrations in terrestrial archives have also been utilised to evaluate TSI on the millennial timescale \citep{Steinhilber09, Steinhilber12, DelaygueBard11, Shapiro11, Vieira11}. Of these, only \citet{Shapiro11} have also considered SSI, while only \citet{Vieira11} have used physics-based models to describe the whole chain of the involved processes at each step. Other models have relied on the linear regressions between the measured irradiance and the input proxies, while \citet{Vieira11} have demonstrated that the linear relationship is actually not applicable. Here we follow the approach of \citet{Vieira11} to update their TSI reconstruction and to also, for the first time, reconstruct SSI from the UV to the far-IR, considering physical process instead of relying on linear relationships. Also, for the first time, we present a reconstruction based on composite \isotope[10][]Be and \isotope[14][]C records.

We use the SATIRE model \citep[Spectral and Total Irradiance REconstruction,][]{Fligge2000, Krivova03, Krivova07, Krivova10, Wenzler04, Wenzler06, Vieira11, Ball14, Yeo14, DasiEspuig14, DasiEspuig16}, which is a family of models based on the assumption that the irradiance variation on timescales longer than about one day is driven solely by changes in the photospheric magnetic field. This model family attributes the irradiance changes to the competing contributions of dark (sunspot and pores) and bright (faculae and the network) surface magnetic features, as all other variable irradiance models do. The brightness of individual components is assumed to be time-independent, but their surface coverage changes with time. The positions and the areas of the various surface magnetic components are most accurately extracted from solar full-disc magnetograms, as used in the SATIRE-S model \citep[``S'' stands for satellite era,][]{Fligge2000, Krivova03, Wenzler06, Ball14, Yeo14}, but these are only available (with a suitable quality and cadence) from 1974 onwards. For the period 1610 -- 1974, SATIRE relies on sunspot observations and the corresponding version is termed SATIRE-T \citep[where ``T'' stands for telescope era,][]{Balmaceda07,Krivova07, Krivova10} and SATIRE-T2 for the version using the surface flux transport model to restore the evolution of the solar surface magnetic flux \citep{DasiEspuig14,DasiEspuig16}. The sunspot number has not been recorded prior to 1610 but can be reconstructed from the cosmogenic isotope concentrations, \isotope[14][]C and \isotope[10][]Be retrieved from tree rings and ice cores, respectively \citep{Beer2000a, Solanki04, Usoskin07, Steinhilber09, Steinhilber12, Vieira11, Usoskin17}.

The production rate of cosmogenic isotopes depends on the intensity of galactic cosmic rays (GCRs) that enter Earth's atmosphere. These cosmic rays have to penetrate the heliosphere and the Earth's magnetic field before interacting with nuclei of atmospheric atoms and producing radionuclides that are later stored in terrestrial archives. Therefore, the abundance of isotopes in archives depends not only on the heliospheric magnetic field, but also on the geomagnetic field \citep[shielding effect,][]{Vonmoos06}. As the GCR flux impinging on Earth is modulated by the solar open magnetic flux \citep{Masarik99, Usoskin02, Caballero04}, the concentration of cosmogenic isotopes is an indirect indicator of solar activity. A distinct inverse proportional relationship between GCR flux and solar activity has been observed \citep{Usoskin17}.

Although one can reconstruct the sunspot number from cosmogenic isotope records, these data have a much lower temporal resolution than the daily observed sunspot number: only decadally averaged values are available on the millennial timescale. Therefore, the SATIRE-T model cannot be applied directly. It has been adapted to deal with this complication by \cite{VieiraSolanki10} and \cite{Vieira11}, and the corresponding version of the model was termed SATIRE-M (``M'' for millennia). \citet{VieiraSolanki10} showed that the relationship between the irradiance and quantities that are usually derived from the cosmogenic data is non-linear, and \citet{Vieira11} used this model to reconstruct the TSI, for the first time avoiding a linear relationship. They used the solar open magnetic flux (OMF) derived by \cite{Usoskin07} from the isotope \isotope[14][]C measurements to reconstruct the solar surface magnetic fields over the Holocene for two geomagnetic models. Recently, \cite{Usoskin16_act} reconstructed sunspot number back to 6755 {\footnotesize BC} from an updated geomagnetic model and two cosmogenic isotope series, \isotope[14][]C \citep{Roth13} and \isotope[10][]Be \citep{Yiou97, Muscheler04, Vonmoos06}. Furthermore, \cite{Wu18_composite} combined one global \isotope[14][]C and six local \isotope[10][]Be series into a single state-of-the-art multi-isotope composite record. They also used it to reconstruct the sunspot number over the Holocene.

Here we reconstruct the TSI and SSI using the updated model of \cite{Vieira11} and the most up-to-date sunspot reconstructions by \cite{Usoskin16_act} and Wu et. al (2018). We also extend the model to reconstruct both the total and the spectral irradiance. The concept and the methods of the SATIRE model are introduced in Section \ref{sect:paper2:Model}. The cosmogenic isotope data are described in Section \ref{subsubsect:paper2:cosmogenic data}, and the reconstruction of solar irradiance is presented in Section \ref{sect:paper2:Results}. The results and conclusions are summarised in Section \ref{sect:paper2:summary}.

\section{Model description}
\label{sect:paper2:Model}

\subsection{The general concept of the SATIRE model}
\label{subsect:paper2:satire model}

In the SATIRE models, the solar surface is described by five components: umbra, penumbra, facula, network, and the quiet Sun. The features evolve with time and move across the visible disc as the Sun rotates. Thus the positions of and the fractional disc coverage by the features change. On each given day this information is taken from solar observations. In particular, when daily high-resolution full-disc solar magnetograms and continuum images are available for the last four decades, the SATIRE-S model is able to accurately replicate the directly measured irradiance variability. However, one has to rely on less complete sets of observations for going further back in time.

The longest record of direct solar observations is the sunspot number that extends back to the Maunder minimum, although with progressively degrading quality (see Sect. \ref{subsubsect:paper2:satiret_mf}). SATIRE-T employs the sunspot numbers to first reconstruct the evolution of the solar photospheric magnetic fields through a set of ordinary differential equations \citep{Solanki2000, Solanki02, VieiraSolanki10}, which is then used as input to the reconstruction of the solar irradiance \citep{Balmaceda07, Krivova07, Krivova10}. For time periods before 1610 only indirect data are available, and therefore \citet{VieiraSolanki10} and \citet{Vieira11} adapted the SATIRE model to use the radioisotope data as input (SATIRE-M). This required an additional step of first reconstructing the solar activity (either the sunspot number or the solar OMF) from the concentrations of \isotope[14][]C or \isotope[10][]Be records \citep[e.g.][]{Solanki04, Steinhilber12, Usoskin14}. In this study, to reconstruct the solar total and spectral irradiance on both centennial and millennial timescales, we use the SATIRE-T model for the period from the Maunder minimum to the present and the SATIRE-M model over the period 6755 {\footnotesize BC} to 1895 {\footnotesize AD}. More details of the SATIRE-T and SATIRE-M models are given in Sect. \ref{subsect:paper2:model_satiret} and Sect. \ref{subsect:paper2:model_satirem}, respectively.

In all versions of the model, the input data are used to recover the fractional solar disc coverage by a given photospheric component at a given time, which is described by the corresponding filling factor $\alpha$ (we note that the definition of the filling factors is somewhat different in the version of the model that uses spatially resolved solar magnetograms and images, but this is irrelevant for this work). The brightness of each component $i$ is time-independent and is calculated following \citet{Unruh99} using the ATLAS9 spectral synthesis code \citep{Kurucz93} from the corresponding solar model atmospheres \citep[for details see][]{Unruh99}. Intensity spectra $I_i(\lambda,\mu)$ ($\mu=\mathrm{cos}\theta$, where $\theta$ is the heliospheric angle) of the quiet Sun, umbra, and penumbra are calculated from the model atmospheres of \cite{Kurucz93, Kurucz05} with effective temperatures of 5777 K, 4500 K, and 5400 K, respectively. The model atmosphere used for faculae and network was adapted from the model `P' by \cite{Fontenla99}. The intensity spectra, $I_i(\lambda, \mu)$, are then integrated over the whole disc (or over the activity belts; see Sect. \ref{subsubsect:paper2:satiret_ff}) to give the disc-integrated brightness spectra, $F_i(\lambda)$, of each component.

The solar irradiance at a given wavelength, $\lambda$ (SSI), at a given time, $t$, is obtained by summing up spectra of the individual photospheric components weighted by their surface coverage, given by their filling factors $\alpha_i$:
\begin{equation}
\label{eq:paper2:SSI_concept}
F(\lambda,t)=\sum\limits_{i= \mathrm{u,p,f,n,q}} \alpha_i(t) F_ i(\lambda).
\end{equation}

The ATLAS9 spectral synthesis code uses the assumption of local thermodynamic equilibrium (LTE), which breaks down in the solar chromosphere where the important parts of solar UV radiation are formed. \citet{Krivova06} and \citet{Yeo14} have shown that the SATIRE spectra need a correction below 300 nm. Following \citet{Yeo14}, between 180 and 300 nm we introduce an offset in the absolute values to match the whole hemispheric interval (WHI) reference solar spectrum \citep{Woods09}, while in the range 115 -- 180 nm, the variability is rescaled with empirical factors derived using the SORCE/SOLSTICE measurements \citep{Snow05}. We note that these corrections do not affect the long-term trend. Finally, TSI is calculated as an integral of SSI over the entire spectral range 115 -- 160\,000 nm.

\subsection{SATIRE-T}
\label{subsect:paper2:model_satiret}

\subsubsection{Surface magnetic field}
\label{subsubsect:paper2:satiret_mf}

In the SATIRE-T model, the evolution of the solar surface magnetic field is described by a set of coupled ordinary differential equations (ODEs). The model has been described in detail in earlier papers \citep{Solanki02, Krivova07, VieiraSolanki10}. Therefore, here we only briefly outline its main features. Three components of the magnetic field are considered: the magnetic field in active regions (ARs;  $\phi_\mathrm{act}$), in ephemeral regions (ERs; $\phi_\mathrm{eph}$), and in the open flux, $\phi_\mathrm{open}$. Active regions include sunspots and faculae. They typically emerge in the so-called activity belts (i.e. within  latitudes of approximately 5 -- 30$\degr$ for sunspots and 5 -- 45$\degr$ for faculae) and show a pronounced solar activity cycle.  Ephemeral regions are small-scale bipolar regions with much shorter lifetimes distributed more homogeneously over the solar surface. They show a much weaker, if any, variation over the solar cycle \citep[see][]{Harvey93, Hagenaar01, Hagenaar03}. The emergence of ARs and ERs in our model is described by the corresponding emergence rates, $\varepsilon_\mathrm{act}$ and $\varepsilon_\mathrm{eph}$, respectively. This magnetic field decays on timescales of $\tau_\mathrm{act}^0$ and $\tau_\mathrm{eph}^0$. Part of the field is trapped by the solar wind plasma and dragged further into the heliosphere, forming the so-called open flux. Whereas part of the open flux decays within a few months, some part of it can survive much longer \citep[about 3 -- 4 years,][]{Solanki2000, Solanki02, VieiraSolanki10, Owens12_HCS}. Thus the slowly  and rapidly decaying components of the open flux, $\phi_\mathrm{open}^\mathrm{s}$ and $\phi_\mathrm{open}^\mathrm{r}$, respectively, are modelled separately, with the corresponding decay times being $\tau_\mathrm{open}^\mathrm{s}$ and $\tau_\mathrm{open}^\mathrm{r}$. The flux transfer from ARs and ERs to the slowly evolving open flux occurs on timescales $\tau_\mathrm{act}^\mathrm{s}$ and $\tau_\mathrm{eph}^\mathrm{s}$, while $\tau_\mathrm{act}^\mathrm{r}$ is the timescale for the transfer of the AR flux to the rapidly evolving open flux. The evolution of the magnetic field is then described by the following equations:
\begin{equation}
\label{eq:paper2:phiact}
\frac{d\phi_\mathrm{act}}{dt}=\varepsilon_\mathrm{act}-\frac{\phi_\mathrm{act}}{\tau_\mathrm{act}^0}-\frac{\phi_\mathrm{act}}{\tau_\mathrm{act}^\mathrm{s}}-\frac{\phi_\mathrm{act}}{\tau_\mathrm{act}^\mathrm{r}},
\end{equation}
\begin{equation}
\frac{d\phi_\mathrm{eph}}{dt}=\varepsilon_\mathrm{eph}-\frac{\phi_\mathrm{eph}}{\tau_\mathrm{eph}^0}-\frac{\phi_\mathrm{eph}}{\tau_\mathrm{eph}^\mathrm{s}},
\end{equation}
\begin{equation}
\frac{d\phi_\mathrm{open}^\mathrm{r}}{dt}=\frac{\phi_\mathrm{act}}{\tau_\mathrm{act}^\mathrm{r}}-\frac{\phi_\mathrm{open}^\mathrm{r}}{\tau_\mathrm{open}^\mathrm{r}},
\end{equation}
\begin{equation}
\label{eq:paper2:phiofs}
\frac{d\phi_\mathrm{open}^\mathrm{s}}{dt}=\frac{\phi_\mathrm{act}}{\tau_\mathrm{act}^\mathrm{s}}+\frac{\phi_\mathrm{eph}}{\tau_\mathrm{eph}^\mathrm{s}}-\frac{\phi_\mathrm{open}^\mathrm{s}}{\tau_\mathrm{open}^\mathrm{s}}.
\end{equation}

As ARs typically include sunspots, the sunspot number can be considered as a good indicator of the emergence of the magnetic field in ARs. \citet{Balmaceda07}, \citet{VieiraSolanki10}, and \citet{Krivova10} used the group sunspot number (GSN) by \citet{HoytSchatten98} as input, while \citet{Krivova07} considered both the group and the International (or Z\"urich) number. The group and the International sunspot numbers agree over the last 150 years but show different overall activity levels between the years 1700 and 1850. A recalibration and multiple corrections to sunspot observations is the subject of an intense ongoing debate \citep{Clette14, Lockwood14_ssn2, Usoskin16_ssn, Cliver16, Chatzistergos17, Svalgaard17}, which remains a controversial topic. The effect of the choice of the input sunspot number on the irradiance reconstructions has been considered by \citet{Kopp16} and is beyond the scope of this study. Briefly, the choice of the input series affects the overall irradiance level in the eighteenth and early nineteenth centuries but not the magnitude of the secular change between the Maunder minimum (1637 -- 1715) and now \citep{Krivova07, Kopp16}. Furthermore, as the main goal of this work is a reconstruction of the solar irradiance over a much longer period using the cosmogenic isotope data, the level of the solar activity in the eighteenth century is derived independently of the sunspot data towards the end of this article.

Here we therefore use the updated International (Z\"urich) sunspot number series (SN version 2.0, SN$_{\rm{v}2}$ hereafter) by WDC-SILSO\footnote{\url{http://sidc.oma.be/silso/datafiles}} as our input sunspot number. However, this SN series only extends back to 1700 which is the end phase of the Maunder minimum. Furthermore, this data set is problematic during the early eighteenth century due to the large amount of missing sunspot number data and the difficulty in calibrating between different observatories. \cite{Vaquero15} have accounted for this issue and revised the sunspot number during this period. We therefore replace the sunspot number between the years 1639 and 1715 with the one reconstructed by \cite{Vaquero15}. The original SN$_{\rm{v}2}$ series is given at three cadences, covering different periods: daily (1818 -- present), monthly (1749 -- present), and yearly (1700 -- present). In this study, we interpolate the monthly sunspot number from 1749 to 1818 and the yearly one averaged before 1749 to a daily cadence for consistency in the input temporal resolution for the SATIRE-T model.

Following \cite{Solanki02}, \cite{Krivova07}, \cite{Krivova10}, and \cite{VieiraSolanki10} we assume the emergence rate of the AR magnetic flux, $\varepsilon_\mathrm{act}$, to be directly related to SN$_{\rm{v}2}$:
\begin{equation}
\varepsilon_\mathrm{act} = \varepsilon^{\mathrm{max},21}\frac{\mathrm{SN_{v2}}}{\mathrm{SN_{v2}^{\mathrm{max},21}}},
\end{equation}
where $\varepsilon^{\mathrm{max},21}$ and $\mathrm{SN_{v2}^{\mathrm{max},21}}$ are the three-month averaged emergence rate and SN value observed during the maximum of cycle 21 \citep{SchrijverHarvey94}, respectively. We take the value of cycle 21 as this one is most comprehensively studied.

We further assume that the emergence rate of ERs is linked to that of the ARs. This is based on the studies of ER evolution by \citet{HarveyMartin73} and \citet{Harvey92, Harvey93}. In particular, \citet{Wilson88} and \citet{Harvey92} found that ERs start emerging earlier than ARs belonging to the same activity cycles. Thus the ER cycles last longer than the corresponding sunspot cycles and overlap during solar minima. Following \cite{VieiraSolanki10}, we define the ER emergence rate ($\varepsilon_\mathrm{eph}$) as follows.
\begin{equation}
\label{eq:paper2:emerg_eph}
\varepsilon_\mathrm{eph}(t)=\sum_{n=1} \varepsilon^{\mathrm{max},n}_\mathrm{act}Xg^n, 
\end{equation}
where $X$ is the amplitude factor and $g^n$ defines the ER cycle shape:
\begin{equation}
  g_{n}(t)= 
  \begin{cases}
  \mathrm{cos}^2 \bigg( \frac{\pi(t-t^\mathrm{max}_n)}{L^\mathrm{eph}_n} \bigg), &-\frac{L^\mathrm{act}_n}{2}-L^\mathrm{ext}_n \leq t-t^\mathrm{max}_n \leq \frac{L^\mathrm{act}_n}{2}+L^\mathrm{ext}_n\\
   0,  &\mathrm{otherwise}.
  \end{cases}
\end{equation}

Here $t^\mathrm{max}_n$ stands for the time when cycle $n$ reaches its maximum and $L^\mathrm{act}_n$ is the corresponding cycle length. The function $g^n$ ensures that the ER cycle maxima are at or before the corresponding sunspot cycle maxima \citep[see][]{DasiEspuig14}. The ER cycle length of the corresponding activity cycle is defined as:
\begin{equation}
L^\mathrm{eph}_n=L^\mathrm{act}_n + 2L^\mathrm{ext}_n,
\end{equation}
where $2L^\mathrm{ext}_n$ is the time extension related to each activity cycle with a constant extension parameter $c_\mathrm{x}$:
\begin{equation}
L^\mathrm{ext}_n=L^\mathrm{act}_n -c_\mathrm{x}.
\end{equation}

The decay and transfer times for the surface magnetic components are free parameters in the SATIRE-T model, which are obtained from a comparison of the results with various observations; details are described in Sect. \ref{subsubsect:paper2:optimisation}.

\subsubsection{Filling factors}
\label{subsubsect:paper2:satiret_ff}

The calculated magnetic flux for each component as a function of time is converted to filling factors entering Eq. (\ref{eq:paper2:SSI_concept}) as follows. First, the fractional disc area coverage by sunspots is obtained directly from the measured sunspot areas whenever these are available \citep{Balmaceda05, Balmaceda09, Yeo17} \footnote{\url{http://www2.mps.mpg.de/projects/sun-climate/data/sunspot_area_psi_20170531.txt}}. 

For the period prior to Greenwich observations (i.e. before 1874) when the area record is not available, we extrapolate this latter coverage using sunspot numbers. Sunspots are divided into umbra and penumbra using their observed mean area ratio: ${\alpha_\mathrm{u}}/{(\alpha_\mathrm{u}+\alpha_\mathrm{p})}={\alpha_\mathrm{u}}/{\alpha_\mathrm{s}}=0.2$ \citep[see, e.g.][]{Solanki03}. Assuming the mean field strength in penumbra is 550 G and in umbra is 1800 G \citep{Keppens96}, we obtain the total magnetic flux in sunspots. The magnetic flux in faculae is then the difference between the total flux in ARs and that in sunspots, $\phi_\mathrm{f} = \phi_\mathrm{act} - \phi_\mathrm{u} - \phi_\mathrm{p}$. The magnetic flux in the network is the sum of the ERs and open flux: $\phi_\mathrm{n} = \phi_\mathrm{eph}+\phi_\mathrm{open}$.

To convert the magnetic flux in faculae $\phi_\mathrm{f}$ and network $\phi_\mathrm{n}$ into filling factors $\alpha_\mathrm{f,n}$, we assume the filling factors to vary linearly with the corresponding magnetic flux until the saturation flux is reached. The filling factors have a value of 1 above the saturation limit \citep{Fligge2000, Krivova03, Krivova07, Yeo14}. The saturation flux in the network ($B_\mathrm{sat,n}$) is fixed at 800 G \citep[for details, see][]{Krivova07, Krivova10}, while the saturation limit in faculae ($B_\mathrm{sat,f}$) is a free parameter in the model \citep{Krivova03}

The derived facular and network magnetic fluxes are disc-integrated, and so this approach assumes that they are homogeneously distributed on the solar surface. This assumption might overestimate the contribution of faculae to the total solar irradiance since in reality, AR faculae may emerge within the so-called activity belts (at latitudes $\pm$5 -- 45$\degr$ for faculae and $\pm$5 -- 30$\degr$ for sunspots), while the brightness of faculae increases toward the limb \citep{Ortiz02, Yeo13}. \cite{Krivova07} used simple weighting factors to account for this. In this study, instead of integrating the intensity spectra, $I_i(\lambda,\mu)$ of sunspots, and ARs over the whole disc, we integrate them over the activity belts.


\begin{table*}
\begin{center}
\caption{Parameters and their allowed ranges in SATIRE-T model.}
\label{table:paper2:satiret_parameters} 
\begin{tabular}{l c r r r r}
\hline\hline  
Parameter \tablefootmark{a}& Notation & \multicolumn{2}{c}{Range} & Kea10 & This work \\
\hline
ER decay time & $\tau_\mathrm{eph}^0$ & & & \multicolumn{2}{c}{0.0016 (fixed)} \\
AR decay time & $\tau_\mathrm{act}^0$ &  &  & 0.3 & 0.3 (fixed) \\
Slow OF decay time & $\tau_\mathrm{open}^\mathrm{s}$ & 0.0016 & 4.0 & 2.97 & 3.75 \\
Rapid OF decay time & $\tau_\mathrm{open}^\mathrm{r}$ & 0.08 & 0.36 & 0.16 & 0.14 \\
AR to slow OF transfer time & $\tau_\mathrm{act}^\mathrm{s}$ & 10 & 90 & 71.2 & 88.3 \\
AR to rapid OF transfer time & $\tau_\mathrm{act}^\mathrm{r}$ & 0.0016 & 3.0 & 2.1 & 2.6 \\
ER to slow OF transfer time & $\tau_\mathrm{eph}^\mathrm{s}$ & 10 & 90 & 17.8 & 20.6 \\
\hline
Saturation flux in faculae & $B_\mathrm{sat,f}$ & 50 & 850 & 156 & 371 \\
Saturation flux in network & $B_\mathrm{sat,n}$ & & & \multicolumn{2}{c}{800 (fixed)} \\
\hline
ER cycle amplitude factor & $X$ & 70 & 150 & 78 & 106 \\
ER cycle extension & $c_\mathrm{x}$ & 5 & 8 & 5.01 & 7.63\\
Fraction of detectable ER flux & $c_\mathrm{e}$ & & & 0.3 (fixed) & 0.4(fixed) \\
\hline 
\end{tabular}
\end{center} 
\tablefoot{
\tablefoottext{a}{Time is in unit [year] and magnetic flux in unit [G].}
}
\tablebib{Kea10: \citet{Krivova10}}.
\end{table*}

\subsubsection{Model optimisation}
\label{subsubsect:paper2:optimisation}

As mentioned in Sect. \ref{subsubsect:paper2:satiret_mf}, some of the decay and transfer times in Eqs. (\ref{eq:paper2:phiact}) -- (\ref{eq:paper2:phiofs}) are free parameters in the SATIRE-T model. There are 12 parameters in total in the SATIRE-T model, as summarised in Table \ref{table:paper2:satiret_parameters}, of which 4 can be inferred from observations whereas others are rather uncertain and are therefore considered to be free. The ER decay time ($\tau_\mathrm{eph}^0$) is taken to be 0.0016 years following \citet{Hagenaar01}, the AR decay time ($\tau_\mathrm{act}^0$) is set to be 0.3 years \citep{Krivova10}, and the saturation field in the network ($B_\mathrm{sat,n}$) is fixed at 800 G \citep{Krivova10}. One more parameter, the fraction of the detectable ER flux, $c_\mathrm{e}$, which is described later in this section is also fixed to 0.4. The other eight parameters are left free within the ranges based on observations and physical constrains \citep[for details, see][]{Krivova07, Krivova10, VieiraSolanki10}. These eight free parameters are fixed from a comparison of the model outcomes with various available observations, as described in the following.

The observations used to compare with and to fix the free parameters are: (1) PMOD TSI composite since 1978 \citep{Froehlich06, Froehlich09}\footnote{\url{http://www.pmodwrc.ch/pmod.php?topic=tsi/composite/SolarConstant}; \\ version d42\_65\_1608}; (2) ground-based total magnetic flux (TMF) measurements by Wilcox Solar Observatory (WSO), National Solar Observatory(NSO) and Mount Wilson Observatory (MWO) covering cycles 21 -- 23 \citep{Arge02, Wang05}; (3) OMF over the period 1845 -- 2010 reconstructed by \cite{Lockwood14_geo} from the $aa$-index; (4) 200 -- 400 nm SSI UV reconstructed from magnetograms with the SATIRE-S model over the period 1976 -- 2015 \citep{Yeo14, Yeo15} and from the f10.7cm radio flux with the EMPIRE model over 1947 -- 1976 \citep{Yeo17}; and (5) facular contribution to the TSI variation calculated with SATIRE-S over 1974 -- 2015 \citep{Yeo14, Yeo15}.

All data sets have been updated since the previous reconstructions by \citet{Krivova07, Krivova10}. A major update concerns the OMF. The updated series spans 60 years longer than the previous version by \cite{Lockwood09} that only covered the period from 1904 to 2010. We note that being the longest observational series, the OMF record mostly affects the long-term trend.  

When comparing the reconstructed TMF to the measured one, we take into account the finding by \citet{KrivovaSolanki04} that more than half of the ER flux is not detected in the magnetogram archives used here due to the insufficient spatial resolution of the instruments or is hidden in the noise. Therefore, when comparing to the observations, we reduce the amount of the ER flux by a factor $c_\mathrm{e}$, therefore the TMF is computed as $\phi_\mathrm{tot}=\phi_\mathrm{act}+c_\mathrm{e}\phi_\mathrm{eph}+\phi_\mathrm{open}$, where $c_\mathrm{e}$ describes the fraction of detectable ER flux. We follow previous studies and take $c_\mathrm{e} = 0.4$ \citep{Balmaceda07, Krivova07, DasiEspuig16}. This factor is not used when calculating the irradiance, as in this case it is the total amount of the solar surface magnetic flux which is of relevance.

To optimise and constrain the free parameters, we utilise the genetic algorithm PIKAIA \citep{Charbonneau95}, which is designed to find the set of parameters that minimizes the difference between the modelled and the reference data sets. We minimize the sum of the reduced chi-squared value ($\chi^2$), which takes the errors of the observations and the number of data points into account. In other words, we search for the maximum of $1/( \chi_\mathrm{TSI}^2+ \chi_\mathrm{faculae}^2 + \chi_\mathrm{UV}^2 + \chi_\mathrm{TMF}^2+ \chi_\mathrm{OMF}^2 )$ \citep[see][for further details]{VieiraSolanki10}. In this way we ensure that no single data set dominates in constraining the free parameters of the model.

\subsection{SATIRE-M}
\label{subsect:paper2:model_satirem}

As direct solar observations are not available before the Maunder minimum, to reconstruct the irradiance before that we use the concentrations of the cosmogenic isotopes \isotope[14][]C and \isotope[10][]Be in natural archives to describe the changes of solar activity with time.

\subsubsection{Cosmogenic data}
\label{subsubsect:paper2:cosmogenic data}

The cosmogenic radioisotopes \isotope[14][]C and \isotope[10][]Be are produced in the terrestrial atmosphere by galactic cosmic rays (GCRs). After entering the Earth's atmosphere, GCRs collide with nuclei (mainly oxygen and nitrogen) and produce these radionuclides. Since the GCR flux is modulated by the heliospheric magnetic field \citep[e.g.][]{Beer2000a, Muscheler04, Usoskin06, Steinhilber08, Vieira11}, this allows a reconstruction of the solar magnetic activity at the time of the isotope production \citep{Lal67}.

To convert the concentrations of the cosmogenic isotopes into Sun-related quantities, several models have to be applied, such as the isotope production model \citep{Masarik99, Kovaltsov12, Poluianov16}, atmospheric transport and deposition model \citep[e.g.,][]{Heikkila09, Roth13}, and the geomagnetic field model \citep{Yang2000, Korte05, Knudsen08, Usoskin16_act}. 

To account for the geomagnetic field, two different models have been used: VDM \citep[virtual dipole moment,][]{Korte05} and VADM \citep[virtual axial dipole moment,][]{Usoskin16_act}. The former is approximated as a geocentric and tilted dipole and is only reliable for up to a few thousand years when the spread and quality of geomagnetic samples are good. The latter, VADM, assumes the dipole to be aligned with the geographical axis and provides a rough geomagnetic estimation for the Holocene. 

After production, \isotope[14][]C takes part in the global carbon circulation in the atmosphere until it is absorbed by plants or stored in sediments \citep{Roth13}, while \isotope[10][]Be attaches to aerosols and quickly precipitates within 1 -- 2 years after the production and is preserved in reservoirs or ice sheets \citep{Vonmoos06}. Unlike \isotope[14][]C, which carries globally averaged information, \isotope[10][]Be records are strongly region-dependent \citep{Sukhodolov17}. Therefore, the signals retrieved from various natural archives are affected by the geochemical process \citep{Steinhilber08, Steinhilber12, Beer12}, or the so-called systematic differences (e.g. snow accumulation rate, long-term changes in the carbon cycle, dating uncertainties).

In this study, three series of decadal SN reconstructed from cosmogenic isotopes are used, as described below. The first, \isotope[14][]C-based  SN series \citep[``U16-14'',][]{Usoskin16_act} is an extended and updated version of a series taken from \cite{Usoskin14}. It is calculated from an updated \isotope[14][]C production rate record INTCAL09 \citep{Reimer09} and a new geomagnetic model (GMAG.9k) reconstructed by \citet{Usoskin16_act}. Based on the geomagnetic model GMAG.9k, \cite{Usoskin16_act} also reconstructed a SN series (``U16-10Be'') from the longest \isotope[10][]Be data set from the Greenland Ice Core Project \citep[GRIP;][]{Yiou97, Muscheler04, Vonmoos06}.

\begin{figure}
\centering
\resizebox{\hsize}{!}{\includegraphics{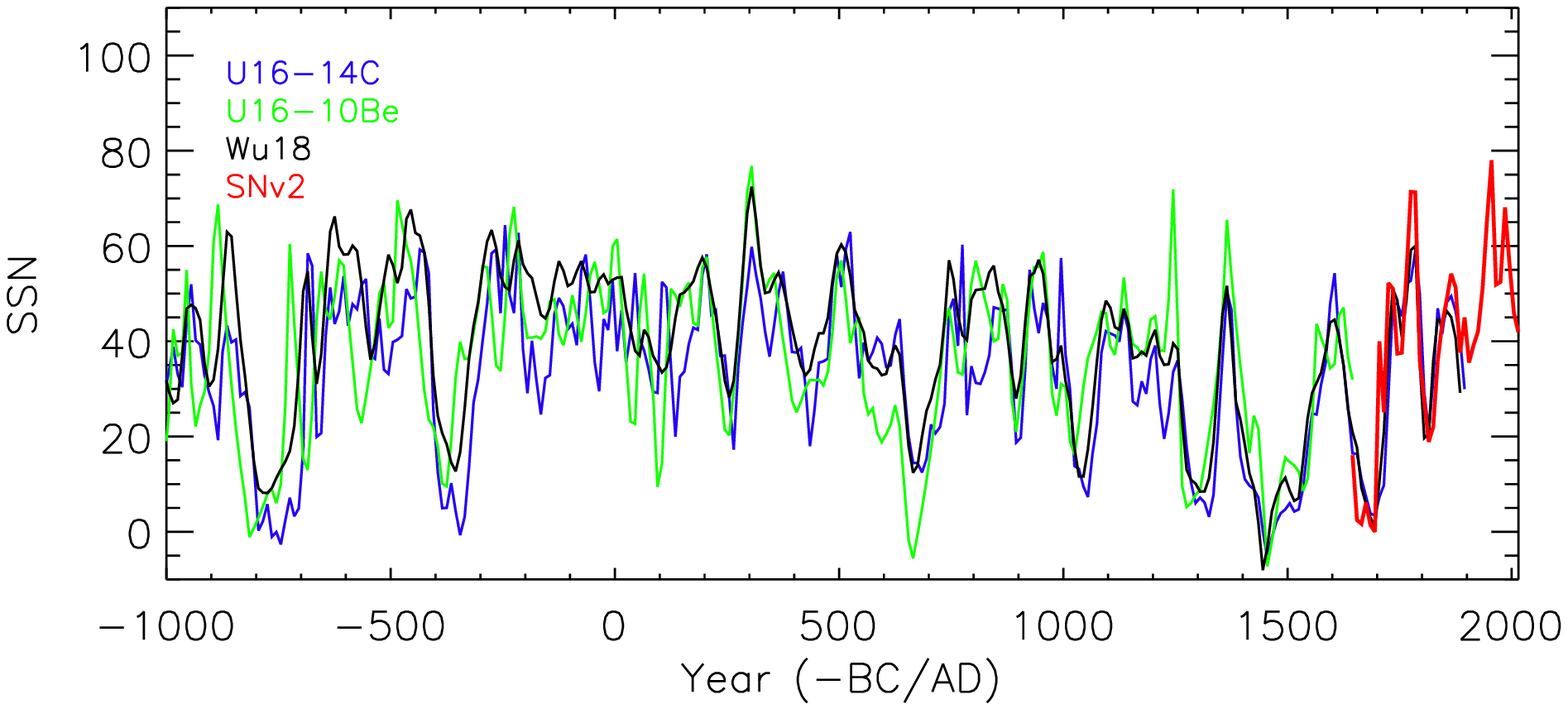}}
\caption{Sunspot series used as input in the SATIRE-M model over the period 1000{\footnotesize BC} to 2015 {\footnotesize AD}. U16-14C (blue): \isotope[14][]C from \citet{Usoskin16_act}; U16-10Be (green): \isotope[10][]Be from \citet{Usoskin16_act}; Wu18 (black): multi-isotope composite constructed by \citet{Wu18_composite}. The decadally averaged directly observed sunspot number SN$_{\rm{v}2}$ (International sunspot number v2.0) used in this paper is shown by a  thick red line for comparison.}
\label{fig:paper2:satirem_ssn}
\end{figure}

Figure \ref{fig:paper2:satirem_ssn} shows the SN series U16-14C (blue) and U16-10Be (green) over the period 1000 {\footnotesize BC} to 2015 {\footnotesize AD}. Due to the different geochemical production processes and non-linear local climatic influences, some temporal discrepancies between multiple isotope data sets can be found (e.g. 900 {\footnotesize BC} -- 700 {\footnotesize BC}). Several studies \citep{DelaygueBard11, Steinhilber12} have combined the individual series using a normalization and the principal component analysis (PCA), which extract the common signals (assumed to be solar) and remove the systematic effects. However, this method keeps only the relative variability, and essentially averages the data points at a given time without taking the geomagnetic fields and the potential time lags into account, and neglects uncertainties. According to \cite{Adolphi16}, the time lags between individual series may reach up to several decades. To account for these issues, \cite{Wu18_composite} combined six \isotope[10][]Be series with the global \isotope[14][]C record into a consistent multi-isotope series, ``Wu18'', the third  series used in this study. This composite takes care of the temporal discrepancies between the seven individual series and minimises the difference between the modelled solar activity and the measured cosmogenic isotope data \cite[see][for more details]{Wu18_composite}, as shown by the thick black curve in Fig. \ref{fig:paper2:satirem_ssn}. For comparison, the decadally averaged directly observed sunspot number series after 1639 is represented by the red solid line. These data sets, their periods, and the geomagnetic field used are summarized in Table \ref{table:paper2:isotope datasets}. 

Until now, the reconstructions of TSI were only based on individual records \citep{Shapiro11, DelaygueBard11, Vieira11}, of which only one reconstruction \citep{Vieira11} was made with a non-linear model. Here we make use of the recent composite, Wu18, to reconstruct the TSI and the first SSI series over the Holocene.

\begin{table*} 
\centering
\caption{Data sets used here, their time spans, and the geomagnetic models.}
\label{table:paper2:isotope datasets}
\begin{tabular}{l c c c | c}
\hline\hline
Date set & Radionuclide & Data spans & SN version & Geomagnetic model\\
\hline
U16-14C & \isotope[14][]C & 6755 {\footnotesize BC} -- 1895 {\footnotesize AD} & U16 & \\
U16-10Be & \isotope[10][]Be & 6755 {\footnotesize BC} -- 1645 {\footnotesize AD} & U16 & VADM; U16 \\
Wu18 & \isotope[14][]C \& \isotope[10][]Be & 6755 {\footnotesize BC} -- 1895 {\footnotesize AD} & Wu18 & \\
\hline
\end{tabular}
\tablebib{U16: \citet{Usoskin16_act}; Wu18: \citet{Wu18_composite}.}
\end{table*}

\subsubsection{Reconstruction of solar irradiance}
\label{subsubsct:papers:satirem_reconstruction of solar irradiance}

The quantity related to solar activity that can be derived most directly from cosmogenic isotope concentrations is the solar modulation potential. It describes the energy loss of GCRs in the heliosphere due to the solar magnetic field \citep{Usoskin05}. The solar modulation potential can then be converted into the solar open magnetic flux (OMF) using the heliospheric cosmic ray transport model \citep{Usoskin07}. 

As described by Eqs. (\ref{eq:paper2:phiact}) -- (\ref{eq:paper2:emerg_eph}), the open magnetic flux is related to ARs and ERs. \cite{VieiraSolanki10} and \cite{Vieira11} made use of this relationship, as well as the other two physical assumptions (see the descriptions below)\ to take the lower temporal resolution of the cosmogenic isotope data into account. This modified model is termed the SATIRE-M model. The first physical assumption is that the fluxes in ARs, ERs, and rapidly evolving open flux are in a steady state on decadal time scale, that is, the emerging flux is approximately equal to the decaying flux \citep{VieiraSolanki10}:
\begin{equation}
\bigg\langle \frac{\phi_\mathrm{act}}{\tau_\mathrm{act}} \bigg\rangle = \langle \varepsilon_\mathrm{act} \rangle,       \bigg\langle \frac{\phi_\mathrm{eph}}{\tau_\mathrm{eph}}\bigg\rangle = \langle \varepsilon_\mathrm{eph} \rangle,      \bigg\langle \frac{\phi_\mathrm{act}}{\tau_\mathrm{act}^r} \bigg\rangle =\bigg \langle \frac{\phi_\mathrm{open}^r}{\tau_\mathrm{open}^r} \bigg\rangle,
\end{equation}
where $\langle ... \rangle$ represents decadal average. The temporal parameters $\tau_\mathrm{act}$ and $\tau_\mathrm{eph}$ are
\begin{equation}
\frac{1}{\tau_\mathrm{act}}=\frac{1}{\tau_\mathrm{act}^0}+\frac{1}{\tau_\mathrm{act}^\mathrm{s}}+\frac{1}{\tau_\mathrm{act}^\mathrm{r}},
\end{equation}
\begin{equation}
\frac{1}{\tau_\mathrm{eph}}=\frac{1}{\tau_\mathrm{eph}^0}+\frac{1}{\tau_\mathrm{eph}^\mathrm{s}}.
\end{equation}

The second physical assumption is that the decadally averaged sunspot number is linearly related to the maximum sunspot number of the corresponding cycle \citep[i.e. $\langle R \rangle \propto R_i^\mathrm{max}$,][]{Usoskin07, VieiraSolanki10}. Based on these assumptions, \cite{VieiraSolanki10} obtained the following relationship between the decadal sunspot number and the solar open flux.
\begin{equation}
\label{eq:paper2:R=aFoj1+bFoj2}
\langle R_\mathrm{i} \rangle_j = a_\mathrm{R} \langle \phi_\mathrm{open} \rangle_j + b_\mathrm{R} \langle \phi_\mathrm{open} \rangle_{j+1},
\end{equation}
where $a_\mathrm{R}=1/c\tau_1$ and $b_\mathrm{R}=1/c\Delta t$, respectively. The sampling interval $\Delta t$ is 10 years, and the constant $c$ is
\begin{equation}
c=\bigg[ \bigg( \frac{1}{\tau_\mathrm{act}^\mathrm{s}}+\frac{\tau_\mathrm{open}^\mathrm{r}}{\tau_\mathrm{open}^\mathrm{s}\tau_\mathrm{act}^\mathrm{r}} \bigg) \tau_\mathrm{act} + \frac{2.2 X \tau_\mathrm{eph}}{\tau_\mathrm{eph}^\mathrm{s}} \bigg] \frac{\varepsilon_\mathrm{act}^{\mathrm{max},21}}{\mathrm{SN_{v2}^{\mathrm{max},21}}},
\end{equation}
and the time parameter $\tau_1$ is
\begin{equation}
\frac{1}{\tau_1}=\frac{1}{\tau_\mathrm{open}^\mathrm{s}}-\frac{1}{\Delta t}.
\end{equation}
The time parameters $\tau_1$, $\tau_\mathrm{act}$, $\tau_\mathrm{eph}$ and the ER amplitude factor $X$ are taken from the SATIRE-T model and are fixed here.

Equation (\ref{eq:paper2:R=aFoj1+bFoj2}) describes the relation between decadally averaged SN and the OMFs from two cycles. Physically, this relationship means that the open flux during a given cycle is affected by solar activity from this and the previous solar cycle. Therefore, the ratio of the contributions from two cycles, $a_\mathrm{R}/b_\mathrm{R} = \Delta t/ \tau_1 = (\Delta t/\tau_\mathrm{open}^\mathrm{s}) -1$, depends on the sampling interval and on the decay time of slowly evolving open flux. In this study, our sampling interval and the $\tau_\mathrm{open}^\mathrm{s}$ are 10 years and 3.75 years, respectively. Thus, the $a_\mathrm{R}/b_\mathrm{R}$ ratio is 1.67, which is lower than the estimate by \cite{Vieira11}, who obtained 2.4. This implies that the contribution of the previous cycle to the open flux at a given time is higher in the updated model.

Next, to calculate the solar irradiance, the filling factors of all magnetic components are needed (see Eq. \ref{eq:paper2:SSI_concept}). We first compute the filling factors for the sunspot umbrae and penumbrae. The relation between the decadal sunspot area (decadally averaged filling factors for umbrae and penumbrae) and the decadal sunspot number can be simplified into a linear relation \citep{Vieira11}:
\begin{equation}
\label{eq:paper2:alphas=A1R+A2}
\langle \alpha_\mathrm{s} \rangle_j =A_1 \langle R_\mathrm{i} \rangle_j +A_2,
\end{equation}
where $A_1$ is a proportionality coefficient and $A_2$ is the offset. Both coefficients are time-independent, as well as the various $a$, $b,$ and $c$ parameters appearing in equations below. Filling factors for umbra and penumbra are obtained by combining Eqs. (\ref{eq:paper2:R=aFoj1+bFoj2}) and (\ref{eq:paper2:alphas=A1R+A2}) and assuming a ratio of umbral to penumbral area of 1:4 \citep{Solanki03}:
\begin{equation}
\label{eq:paper2:alphau}
\langle \alpha_\mathrm{u} \rangle_j =  a_\mathrm{u} \langle \phi_\mathrm{open} \rangle_j + b_\mathrm{u} \langle \phi_\mathrm{open} \rangle_{j+1} + c_\mathrm{u},
\end{equation}
\begin{equation}
\langle \alpha_\mathrm{p} \rangle_j =  a_\mathrm{p} \langle \phi_\mathrm{open} \rangle_j + b_\mathrm{p} \langle \phi_\mathrm{open} \rangle_{j+1} + c_\mathrm{p}.
\end{equation}
The values of $a_\mathrm{u,p}$, $b_\mathrm{u,p}$, and $c_\mathrm{u,p}$ used in this study are listed in Table \ref{table:paper2:satirem_parameters}.

The filling factors for faculae and network are found similarly to SATIRE-T while taking the decadal average into account. Thus, we assume (1) the mean field strength of umbra ([$B_\mathrm{u}$]) and penumbra ([$B_\mathrm{p}$]) to be 1800 G and 550 G, respectively; (2) an AR is composed of sunspot and faculae; and (3) the network includes ERs and OMF, that is, $\phi_\mathrm{n} = \phi_\mathrm{eph}+\phi_\mathrm{open}$ \citep{Krivova10}. The filling factors of faculae and network can then be written as
\begin{equation}
\langle \alpha_\mathrm{f} \rangle_j = \frac{\langle \phi_\mathrm{f} \rangle_j}{S_\sun B_\mathrm{sat,f}} =a_\mathrm{f} \langle \phi_\mathrm{open} \rangle_j + b_\mathrm{f} \langle \phi_\mathrm{open} \rangle_{j+1} + c_\mathrm{f},
\end{equation}
\begin{equation}
\label{eq:paper2:alphan}
\langle \alpha_\mathrm{n} \rangle_j = \frac{\langle \phi_\mathrm{n} \rangle_j}{S_\sun B_\mathrm{sat,n}} = a_\mathrm{n} \langle \phi_\mathrm{open} \rangle_j + b_\mathrm{n} \langle \phi_\mathrm{open} \rangle_{j+1},
\end{equation}
where $S_\sun$ is the solar surface area. The magnetic saturation limits of faculae ($B_\mathrm{sat,f}$) and network ($B_\mathrm{sat,n}$) are adapted from SATIRE-T as well. The values of $a_\mathrm{f,n}$, $b_\mathrm{f,n}$ , and $c_\mathrm{f}$ used in this study are listed in Table \ref{table:paper2:satirem_parameters}.

Finally, the solar spectral irradiance can be obtained by applying these filling factors of the magnetic components in Eq. (\ref{eq:paper2:SSI_concept}):
\begin{equation}
\label{eq:paper2:SSI_satirem}
\langle F(\lambda,t) \rangle_j = a_\mathrm{F}(\lambda) \langle \phi_\mathrm{open} \rangle_j + b_\mathrm{F}(\lambda) \langle \phi_\mathrm{open} \rangle_{j+1} + F_\mathrm{q}(\lambda),
\end{equation}
where
\begin{equation}
(a_\mathrm{F},b_\mathrm{F})(\lambda)= \sum \limits_{i= \mathrm{u,p,f,n}} (a_i,b_i) \langle F_i(\lambda) - F_{\mathrm{q}}(\lambda) \rangle.
\end{equation}

It is important to note that with Eq. (\ref{eq:paper2:SSI_satirem}), the solar spectral and total irradiance can be computed straightforwardly from the values of the open flux from two consecutive cycles. Moreover, during the simplification of the SATIRE-M model in order to deal with the decadal cadence, the model inherits all the parameters from the SATIRE-T model (Sect. \ref{table:paper2:satiret_parameters}) and no free parameters are introduced.

\begin{table}[htb]
\centering
\caption{Parameters used in Eqs. (\ref{eq:paper2:alphau}) -- (\ref{eq:paper2:alphan}).}
\label{table:paper2:satirem_parameters}
\begin{tabular}{c c c c c}
\hline\hline
Quantity & a & b  & c & a/b \\
         & [$10^{-5}$ Wb$^{-1}$] & [$10^{-5}$ Wb$^{-1}$] &  & \\
\hline
$\alpha_\mathrm{u}$ & 3.2  & 1.9 & -0.4 & 1.67 \\
$\alpha_\mathrm{p}$ & 12.8  & 7.7  & -1.6 & 1.67 \\
$\alpha_\mathrm{f}$ & 126.3 & 75.7 & -9.1  & 1.67 \\
$\alpha_\mathrm{n}$ & 71.8  & 30.8 &  -    & 2.33 \\
\hline
\end{tabular}
\end{table}

\section{Results}
\label{sect:paper2:Results}

\subsection{Comparison with observations}

\begin{figure}
\centering
\resizebox{\hsize}{!}{\includegraphics{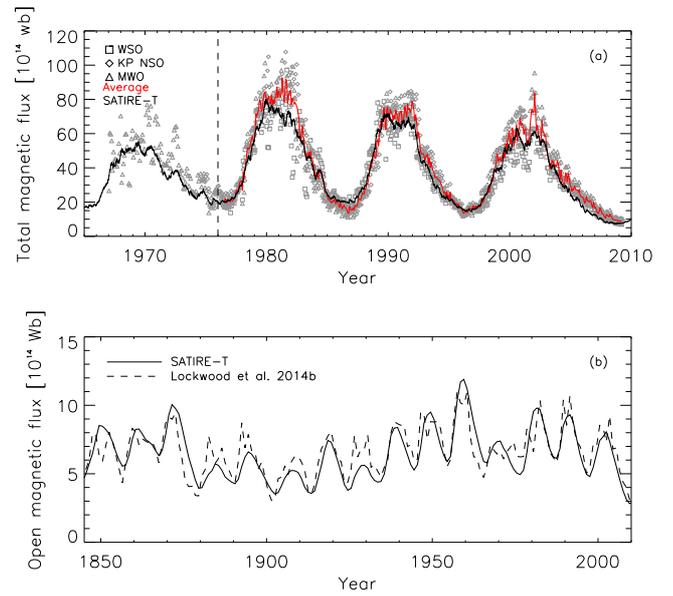}}
\caption{a) Three-month moving averaged reconstructed total magnetic flux (thick black line) compared to the averages of the three observational data sets (thick red line). The three individual observational data sets are WSO (squares), KP NSO (diamonds), and MWO (triangles). The reconstructed total magnetic flux shown is $\phi_\mathrm{tot}=\phi_\mathrm{act}+ 0.4\phi_\mathrm{eph}+\phi_\mathrm{open}$, to take the unresolved magnetic flux in small ephemeral regions into account (see Sect. \ref{subsubsect:paper2:optimisation}). b) Yearly averaged reconstructed OMF (solid line) since 1845 and the reconstruction from the $aa$-index by \citet[][dashed line]{Lockwood14_geo}.}
\label{fig:paper2:satiret_flux_compareobservation}
\end{figure}

\begin{figure}
\centering
\resizebox{\hsize}{!}{\includegraphics{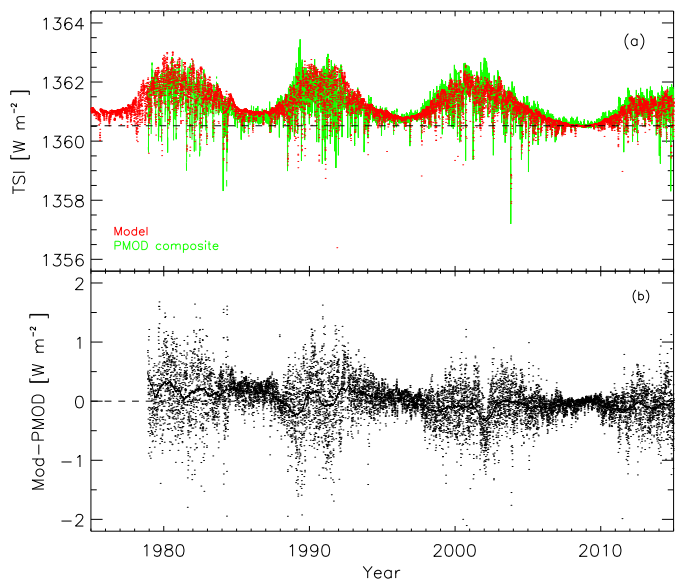}}
\caption{ TSI since 1978: a) Daily reconstructed (red) and measured PMOD composite (green). b) Difference between the reconstruction and the PMOD composite. Daily values are indicated by dots and the 361-day moving average by the thick solid line.}
\label{fig:paper2:satiret_tsi_compareobservation}
\end{figure}

As described in Sect. \ref{subsubsect:paper2:optimisation}, we fix the free parameters of the SATIRE-T model by comparing the output with five independent data records. Here, we first compare the reconstructed quantities to the corresponding observations or alternative models over their overlapping periods of time.

The modelled total magnetic flux (TMF) is compared with the observations in Fig. \ref{fig:paper2:satiret_flux_compareobservation}. Panel (a) shows the TMF over cycles 21 -- 23. Data from different observatories are represented by different symbols: WSO (squares), KP NSO (diamonds), and MWO (triangles). Each individual symbol represents the total photospheric magnetic flux for a given Carrington rotation. The thick solid line is a three-month moving average of the reconstruction. The detectable ER flux is reduced by a factor of 0.4 to take the under-detected flux in observations into account. The figure shows the data since 1967, but we only calculate the reduced $\chi^2$ value from 1976 onwards, i.e. to the right of the vertical dashed line, when all three observational data sets are available. In contrast to the previous work by \citet[][Kea10 hereafter]{Krivova10}, who fitted the modelled TMF to individual observation data points, we fit to the average of the three, as indicated by the red line. The reduced $\chi^2$ value is 0.036. Panel (b) compares the modelled OMF (solid line, yearly averages) with the empirical reconstruction by \citet[][dashed line]{Lockwood14_geo} based on the $aa$-index between 1850 and 2010. The reduced $\chi^2$ value over this period is 0.225. We note that although the reference open magnetic flux record used in this paper covers 60 years longer than the previous version \citep{Lockwood09} used by Kea10, our model has a lower $\chi^2$ value than in Kea10 ($\chi^2$=0.243).

Figure \ref{fig:paper2:satiret_tsi_compareobservation}a shows the TSI reconstruction (red dots) from the sunspot number (SN$_{\rm{v}2}$) overplotted on the PMOD composite (green lines) over the period 1978 -- 2015. In this study, we take the SORCE/TIM measurement during the activity minimum in 2008 (1360.52 W/m$^2$) as reference. Both, the reconstructed TSI and PMOD series are normalized to the 2008 minimum level, indicated by the dashed horizontal line. The difference between the two series is shown in panel (b) (daily values as dots and the 361-day moving average as thick solid line). The reduced $\chi^2$ value between the daily modelled TSI and the PMOD composite is 0.20, which is lower than 0.233 obtained by Kea10.

\begin{figure}
\centering
\resizebox{\hsize}{!}{\includegraphics{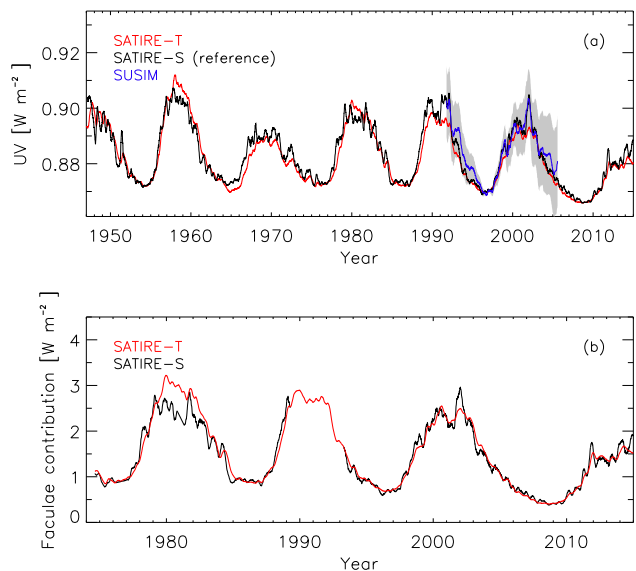}}
\caption{a) The UV irradiance integrated over the range 220 -- 240 nm as reconstructed here (red), modelled with SATIRE-S (black; reference data set from \citet{Yeo14}) and measured by UARS/SUSIM (blue, \citet{Floyd03}). The long-term uncertainty of the UARS/SUSIM measurements (taking the 1996 minimum as a reference) is shown by the shaded area. b) The facular contribution to the TSI changes modelled with SATIRE-T (red) and obtained from full-disc solar magnetograms (SATIRE-S; black). In the SATIRE-S model the period between 1990 and 1993 is not shown due to gaps in the input data \citep{Yeo14}. All curves are in 81-day moving averages.}
\label{fig:paper2:satiret_uv_faculae}
\end{figure}

Next, we compare the modelled UV flux (integral over 220 -- 240 nm; red) with the reference data (black) in Fig. \ref{fig:paper2:satiret_uv_faculae}a. The obtained $\chi^2$ value of 0.043 is significantly lower than the value of 0.072 obtained by Kea10. The reference UV time series here is the reconstruction with SATIRE-S model which has previously been shown to be in close agreement ($R^2_\mathrm{c}$=0.975) with the UARS/SUSIM record \citep[blue line in Fig. \ref{fig:paper2:satiret_uv_faculae}a,][]{Brueckner93, Floyd03}. The long-term uncertainty of the SUSIM instrument is shown by the shaded area, taking the minimum in 1996 as reference \citep{Morrill14, Yeo15}. We take the SATIRE-S reconstruction here rather than the SUSIM data, since it covers a much longer period \citep[40 years vs. 11 years,][]{Krivova06, Yeo14}. The improvement in the UV band is largely due to the more accurate description of the activity belts (see Sect. \ref{subsubsect:paper2:satiret_ff}). By running our model with exactly the same set-up but without accounting for activity belt, we obtain a higher $\chi^2$ value of 0.06.

Since the irradiance variations in the UV range are mainly due to faculae \citep{Unruh08}, we additionally constrain the SATIRE-T model by requiring that it also reproduces the facular contribution to the TSI variation correctly. The calculated facular contribution to the TSI variation over the period 1974 -- 2015 (red) is compared to that obtained from the more accurate SATIRE-S model based on full-disc solar magnetograms \citep[black,][]{Yeo14} in Fig. \ref{fig:paper2:satiret_uv_faculae}b. Since the direct information on the emergence rate of bright features (faculae and network) back to 1610 is missing, their emergence rate is assumed to be related to the evolution of sunspots. This is a valid assumption on timescales of months or longer, but the modelled facular variability cannot be expected to be accurate on timescales shorter than a month. Therefore, the facular contribution is shown as an 81-day moving average. The reduced $\chi^2$ value between the two series is 0.018, which is almost four times lower than in Kea10. It is also worth noting that both the secular change and the solar cycle amplitude of the UV fluxes and the facular contribution to the TSI changes agree very well with the SATIRE-S model.

Please note that the UV and faculae contribution to the TSI are smoothed with an 81-day moving average, which leads to artificially higher correlation coefficients as in this case there are fewer independent points. This is done in order to compare directly with the previous version.

\begin{figure}
\centering
\resizebox{\hsize}{!}{\includegraphics{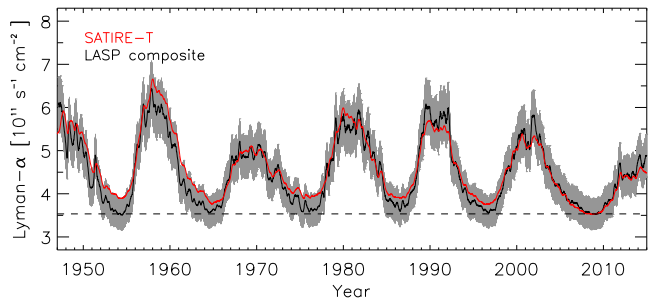}}
\caption{81-day moving average of the reconstructed Lyman-$\alpha$ (red) and the Lyman-$\alpha$ composite by \cite[][see text]{Woods04} with its 1-$\sigma$ uncertainty (grey). The model is normalized to the composite during the 2008 solar minimum (dashed horizontal line).}
\label{fig:paper2:satiret_lyman}
\end{figure}

After constraining the SATIRE-T parameters by comparing to various independent data sets as described above, we also compare the modelled Lyman-$\alpha$ irradiance with the LASP composite of measurements (since 1978) and a proxy model \cite[prior to 1978,][]{Woods2000, Woods04}\footnote{\url{http://lasp.colorado.edu/lisird/lya/}}. The Lyman-$\alpha$ line impacts the chemistry and heating rates in the terrestrial middle atmosphere and is therefore of interest for climate research. A comparison over the period 1947 -- 2015 is shown in Fig. \ref{fig:paper2:satiret_lyman}; the modelled Lyman-$\alpha$ is in red and the composite is in black. The 1-$\sigma$ uncertainty of the composite is indicated in grey. The modelled Lyman-$\alpha$ is normalized to the 2008 minimum (horizontal dashed line). The reduced $\chi^2$ is 0.013. We emphasise that this data set was not used to constrain the model, meaning that the good agreement between the model and the data provides further support to our model. 

In summary, the resulting reconstructions show excellent agreement with the corresponding reference data. The derived free parameters are listed in Table \ref{table:paper2:satiret_parameters}. The reduced $\chi^2$ values and the correlation coefficients ($R_\mathrm{c}$) of the physical quantities are summarised in Table \ref{table:paper2:compare chi2}, compared with the results of Kea10. The parameters are now fixed and further employed to reconstruct the solar spectral irradiance over the last nine millennia (Sect. \ref{subsect:paper2:model_satirem}).

\begin{table*}
\centering 
\caption{The reduced $\chi^2$ values and the correlation coefficients ($R_\mathrm{c}$) between the SATIRE-T reconstruction and the independent data sets used to constrain the model (except the Lyman-$\alpha$, which was not used for fixing the parameters).}
\label{table:paper2:compare chi2}
\begin{tabular}{c c c c c c c}
\hline\hline
Quantity & Cadence & Period & $\chi^2$ & $R_\mathrm{c}$ & $\chi^2$ & $R_\mathrm{c}$ \\
Version & & & \multicolumn{2}{c}{Kea10} & \multicolumn{2}{c}{This work} \\
\hline
TSI & 1 d & 1978 -- 2015 & 0.233 & 0.81 & 0.200 & 0.80 \\
TMF & 1 CR \tablefootmark{a} & cycle 21 -- 23 & 0.069 & 0.93 & 0.036 & 0.98\\
OMF & 1 y & 1845 -- 2010 & 0.248 & 0.86 & 0.243 & 0.82\\
UV (220 -- 240 nm) & 3 m & 1947 -- 2015 & 0.072 & 0.94 & 0.043 & 0.96\\
Facular contribution to TSI variation & 3 m & 1974 -- 2015 & 0.064 & 0.94 & 0.018 & 0.97\\
\hline
Lyman-$\alpha$ & 3 m & 1947 -- 2015 & - & - & 0.13 & 0.96\\
\hline
\end{tabular}
\tablefoot{
\tablefoottext{a}{Carrington rotation.}
}
\end{table*}

\subsection{SATIRE-T reconstruction back to the Maunder minimum}
\label{subsect:paper2:Recon on centennial}

\begin{figure}
\centering
\resizebox{\hsize}{!}{\includegraphics{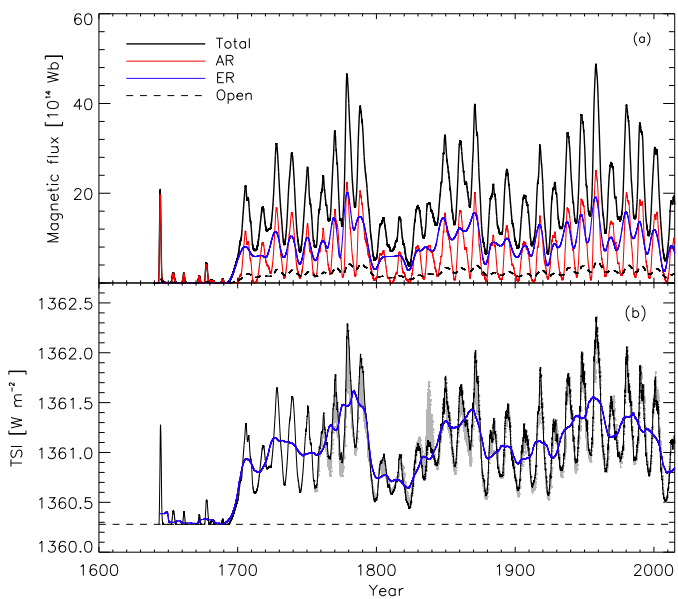}}
\caption{a) Reconstructed magnetic flux (361-day moving averages) at the solar surface since 1639: total magnetic flux (thick black), AR flux (red), ER flux (blue) and open flux (dashed). b) Reconstructed TSI back to 1639 (361-day moving averages in black and 11-year smooth in blue). The horizontal dashed line shows the Maunder minimum level. The gray shaded area represents the uncertainty range due to the choice of the sunspot record (see main text for details).}
\label{fig:paper2:satiret_reconstruction}
\end{figure}

Figure \ref{fig:paper2:satiret_reconstruction}a shows the 361-day smoothed reconstructions of the TMF, AR flux, ER flux, and OMF. We note that the TMF is without the reduction by $c_\mathrm{e}=0.4$, as here we are interested in the total flux responsible for irradiance changes, not the one observed. In our reconstruction, the ER cycles are shorter than the previous versions. It is important to note that while the reconstructed fluxes during the Maunder minimum drop down to zero, this does not imply that the solar activity was completely absent at that time; it is only a consequence of the lack of information on solar activity at that time. Since the sunspot number dropped to zero, the emergence rates of magnetic regions in our model are also zero over most of that period. In principle, the magnetic flux might still have emerged in some weaker regions and contributed to the OMF, which is supported by the cosmogenic data \citep{Stuiver93, Kocharov95, Beer98}. This means that the irradiance drop during the Maunder minimum was probably somewhat less than the value returned by our model, and the estimated secular change is therefore an upper limit.

The reconstructed TSI (SSI integrated over 115 -- 160\,000 nm) is shown in Fig. \ref{fig:paper2:satiret_reconstruction}b (361-day moving average in black, 11-year running mean in blue). The dashed horizontal line indicates the Maunder minimum level of TSI. To compare with other studies, we calculate the increase in the TSI from the end of the Maunder minimum to the mean over the period 1975 -- 2005. Our reconstruction returns an increase of 0.92$^{+0.08}_{-0.02}$ W/m$^2$, which although still within the uncertainty ranges is somewhat lower than in the previous reconstructions with the SATIRE-T model \citep[1.25$^{+0.05}_{-0.3}$ W/m$^2$,][]{Krivova07, Krivova10}, those with SATIRE-T2 based on synthetic magnetograms produced with a surface flux transport model from the group sunspot number \citep[1.2$^{+0.2}_{-0.3}$ W/m$^2$,][]{DasiEspuig16}, and the empirical reconstruction with the NRLTSI model \citep[1.1 W/m$^2$,][]{Coddington16}.

Since the sunspot number is the only input to the SATIRE-T model, the choice of the sunspot number series affects the level in the modelled solar irradiance during the period from the eighteenth to early nineteenth century. However, the secular change is not affected \citep[c.f.,][]{Kopp16_secular}, since neither the modern nor the Maunder minimum levels are affected by this choice. One of the shortcomings of most existing sunspot number reconstructions is the employment of a linear scaling between various individual records, which is not justified \citep{Usoskin17}. Among the proposed revised sunspot number series (Sect. \ref{subsubsect:paper2:satiret_mf}), \cite{Usoskin16_ssn} and \cite{Chatzistergos17} were the only ones to apply a non-linear non-parametric calibration method. \cite{Chatzistergos17} did the calibration using probability distribution functions (PDF) of several backbone observers with errors estimated with Monte Carlo simulations. Therefore we also consider their series here as an alternative input record to the SATIRE-T model. The difference between the TSI reconstructions from the sunspot number records by WDC-SILSO and \cite{Chatzistergos17} is used here as a rough estimate of the uncertainty range over 1739 -- 2010 due to the choice of the sunspot number (grey shading in Fig. \ref{fig:paper2:satiret_reconstruction}b).

To test the stability of our reconstruction and the sensitivity of the model to various free parameters, we have considered different combinations of parameters. In particular, we either fixed $\tau_\mathrm{act}^0$, $B_\mathrm{sat,n}$ , and $c_\mathrm{e}$ as listed in Table \ref{table:paper2:satiret_parameters}, or let them be free parameters within ranges 0.2 -- 0.8, 50 -- 850, and 0.2 -- 0.5, respectively. These tests showed that the resulting reconstructions of TSI differed by up to 0.1 W/m$^2$ on decadal timescales. The overall secular change always lies in the range 0.90 -- 1.0 W/m$^2$. Consequently, we conclude that the reconstruction is independent of the exact choice of the free parameters \citep[see also][]{Krivova07, DasiEspuig16}.

\subsection{SATIRE-M reconstruction on millennial timescales}
\label{subsec:paper2:Recon on millennial}

\begin{figure}
\centering
\resizebox{\hsize}{!}{\includegraphics{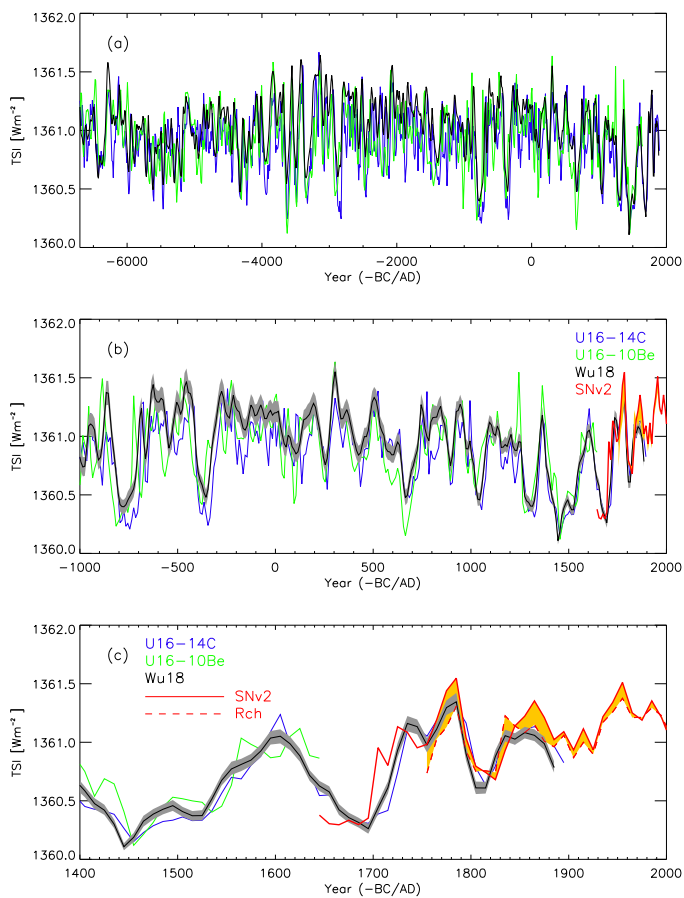}}
\caption{Three TSI reconstructions based on different isotope records: U16-14C (blue), U16-10Be (green) and the composite Wu18 (black) plotted over the periods: a) 6755 {\footnotesize BC} -- 2000 {\footnotesize AD} b) 1000 {\footnotesize BC} -- 2000 {\footnotesize AD} and c) 1400 {\footnotesize AD} -- 2000 {\footnotesize AD}. The uncertainty range of the reconstruction based on the Wu18 composite is indicated by the grey shaded area. Decadal averages of the reconstruction based on the directly observed SN are shown in red after 1639. The solid line is the reconstruction based on the WDC-SILSO series while the dashed line is based on the record by \cite{Chatzistergos17}. The yellow shaded area highlights the difference between them.}
\label{fig:paper2:satirem_tsi}
\end{figure}

Keeping all the parameters fixed, we now employ the SATIRE-M model to reconstruct solar total and spectral irradiance back to 6755 {\footnotesize BC} from the cosmogenic isotope data. Figure \ref{fig:paper2:satirem_tsi}a shows the TSI reconstructions over approximately the last 9000 years based on the three isotope series, as introduced in Sect. \ref{subsect:paper2:model_satirem} (U16-14C in blue, U16-10Be in green, Wu18 in black). Figures \ref{fig:paper2:satirem_tsi}b and \ref{fig:paper2:satirem_tsi}c are enlargements over the periods 1000 {\footnotesize BC} -- 2000 {\footnotesize AD} and 1400 {\footnotesize AD} -- 2000 {\footnotesize AD}, respectively. The decadally averaged TSI reconstruction based on the SN$_{\rm{v}2}$ series is also shown (solid red) for comparison. The reconstruction based on the sunspot record by \cite{Chatzistergos17} is shown by the dashed red curve. Yellow shading highlights the difference between these two reconstructions.

It is worth noticing that the reconstructions based solely on either \isotope[14][]C (U16-14C) or \isotope[10][]Be (U16-10Be) show temporal discrepancies due to differences in the relevant geochemical production processes and local climatic influences \citep[Sect. \ref{subsubsect:paper2:cosmogenic data} and ][]{Wu18_composite}. The reconstruction based on the composite (Wu18) of one global \isotope[14][]C and six regional \isotope[10][]Be series has remedied this \citep[See Fig. 12 in][]{Wu18_composite}, and shows the common solar signal extracted from seven individual isotope series. We therefore consider this latter reconstruction to be more robust than reconstructions based on individual isotope data. On the millennial timescale, the variability of TSI ($\Delta \mathrm{TSI}/ \overline{\mathrm{TSI}}$) obtained from all three isotope records is very similar ($\approx$0.11\%). Since in the second half of the twentieth century, the activity of the Sun was comparatively high \citep[see also][]{Solanki04}, this result means that during extended activity minima (grand minima) the irradiance dropped by up to 1.5 W/m$^2$ compared to the recent level. Interestingly, during the nineteenth century the TSI reconstructed from the cosmogenic isotope data agrees better with the reconstruction from the sunspot record by \citet[][dashed red curve]{Chatzistergos17} than with the one  from the WDC-SILSO (solid red curve). The reduced $\chi^2$ calculated over the period 1750 -- 1900 is 0.32 and 0.72 for the R$_{CH}$ and WDC-SILSO-based reconstructions, respectively.

\begin{figure}
\centering
\resizebox{\hsize}{!}{\includegraphics{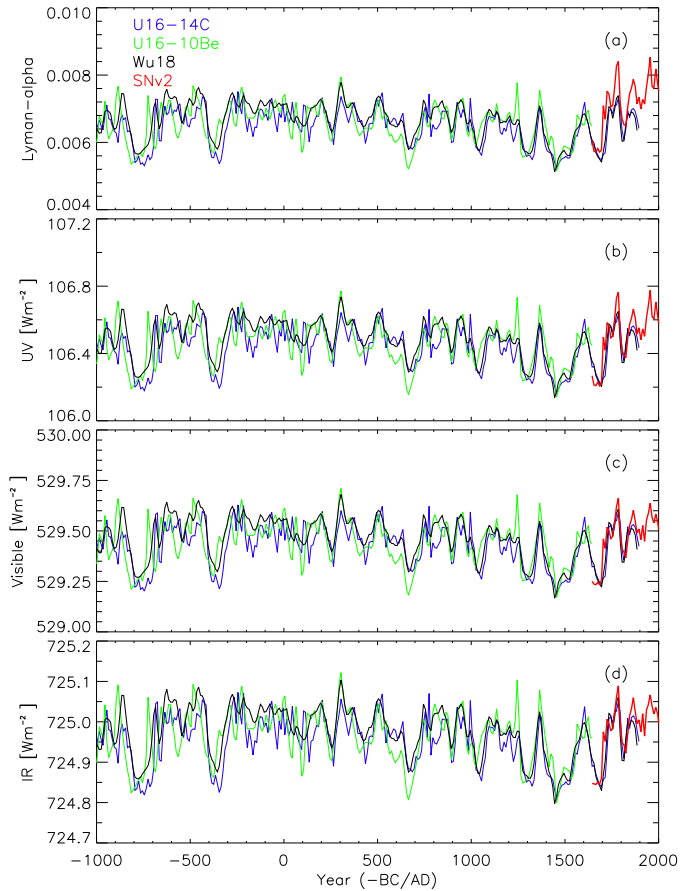}}
\caption{Reconstruction of UV/SSI integrated over a) Lyman-$\alpha$, b) 115 -- 400 nm (marked UV on the axis label), c) 400 -- 700 nm and d) wavelengths longward of 700 nm (IR) based on three isotope-based SN series (see legend in the plot) over the period 1000 {\footnotesize BC} -- 2000 {\footnotesize AD}. The reconstruction based on the directly observed SN is shown in thick red line after 1639. All curves are decadal averages.}
\label{fig:paper2:satirem_ssi}
\end{figure}

Figure \ref{fig:paper2:satirem_ssi} shows the reconstructed SSI in selected wavelength bands over the period 1000 {\footnotesize BC} -- 2000 {\footnotesize AD}. The four selected spectral intervals are (a) Lyman-$\alpha$, here integral over 121 -- 122 nm, (b) UV between 115 and 400 nm, (c) visible between 400 and 700 nm and (d) IR, longwards of 700 nm.

\section{Summary} 
\label{sect:paper2:summary}

Solar radiation is the dominant energy source for the climate of Earth. In addition to the known quasi 11-year solar cycle, the Sun varies on longer timescales, and this variability might affect the Earth. Since direct solar irradiance measurements only exist after 1978, this record is too short to properly assess the solar influence on climate, meaning that reconstructions of past irradiance changes are needed. Models assuming that irradiance variations on timescales longer than about a day are caused by the evolution of the solar surface magnetism reproduce over 90\% of the measured irradiance variability \citep[e.g.][]{Krivova03, Ermolli03, Ermolli13, Ball14, Yeo14, Coddington16}; in particular, the newest version of SATIRE \citep{Yeo17} does so without any recourse to the irradiance measurements (i.e. without free parameters). Therefore, here we assume that the same is also true on longer timescales up to millennia and reconstruct solar total and spectral irradiance over the last 9 millennia from proxies of solar magnetic activity. In particular, we use the SATIRE-T \citep{Krivova07, Krivova10} and -M \citep{Vieira11} models, in which the evolution of the solar surface magnetic field is reconstructed from the sunspot number \citep{Clette14, Clette16, Chatzistergos17} and cosmogenic isotope data \citep{Usoskin16_act, Wu18_composite}, respectively. 

First, the SATIRE-T model has been re-visited with several modifications and updates. The changes include: modified ephemeral region cycle description (to assure the fact that the ER cycle maxima occur at or before the corresponding AR maxima), a more realistic description of the spatial distribution of faculae and sunspots (activity belts), and updates in the input sunspot number series and the reference series (OMF, TMF, TSI, UV and facular contribution to the TSI variation). The reconstruction with the SATIRE-T is in close agreement with the observations over the last four decades. The modelled open magnetic flux also agrees well with the independent reconstruction based on the $aa$-index over the last 150 years \citep{Lockwood14_ssn2}. The reconstructed TSI shows an increase between the late seventeenth century and the present of 0.92$^{+0.15}_{-0.05}$  W/m$^2$, which is somewhat lower (but within the mutual uncertainty ranges) than in the earlier reconstructions with SATIRE-T \citep{Krivova07, Krivova10}, SATIRE-T2 \citep{DasiEspuig16}, and the NRLTSI model \citep{Coddington16}. This is significantly lower than in the reconstruction by \cite{Shapiro10} and \cite{Egorova18}.

Keeping the free parameters fixed, we then used the SATIRE-M model and the SN series reconstructed from cosmogenic data to calculate the TSI and SSI over the last 9000 years. Of the three isotope-based SN series, two are individual \isotope[14][]C and \isotope[10][]Be series from \cite{Usoskin16_act}, and one is the newest multi-isotope composite by \cite{Wu18_composite}. The three TSI/SSI reconstructions with the SATIRE-M model show a mutually similar long-term variability. The range of the TSI variability on millennial timescales for the three used isotope series is about 0.11\% (1.5 W/m$^2$). After the Maunder minimum, the reconstruction from the cosmogenic isotopes is consistent with that from the direct SN observation. Furthermore, over the nineteenth century, the agreement with the reconstruction from the SN by \cite{Chatzistergos17} is better than with that from the WDC-SILSO SN series \citep{Clette14}. 

This reconstruction is the first SSI reconstruction over the Holocene, which uses physics-based models to describe all involved processes \citep{VieiraSolanki10, Vieira11}, and also the first one based on a composite \isotope[14][]C-\isotope[10][]Be record \citep{Wu18_composite}, and is recommended for studies of long-term climate change \citep[e.g.][]{Jungclaus16}. The TSI/SSI reconstructions are available at the web page of Max Planck Institute for Solar System Research ``Solar Variability and Climate'' group\footnote{\url{http://www2.mps.mpg.de/projects/sun-climate/data.html}}.


\begin{acknowledgements}

The authors thank the LASP Interactive Solar Irradiance Data Center for the Lyman-$\alpha$ composite, the WDC-SILSO (Royal Observatory of Belgium, Brussels) for the sunspot number series (\url{http://www.sidc.be/silso/datafiles}). The SUSIM data is available at \url{http://wwwsolar.nrl.navy.mil/uars/}. C.-J. Wu acknowledges support by the Max Planck Research School (IMPRS). I. U. acknowledges support by the Center of Excellence ReSoLVE (Academy of Finland Project No. 272157).
\end{acknowledgements}


\bibliographystyle{aa}
\bibliography{ref1}

\end{document}